\documentstyle[twoside,psfig]{article}

\catcode`\@=11
\long\def\@makefntext#1{
\protect\noindent \hbox to 3.2pt {\hskip-.9pt  
$^{{\eightrm\@thefnmark}}$\hfil}#1\hfill}               

\def\thefootnote{\fnsymbol{footnote}}
\def\@makefnmark{\hbox to 0pt{$^{\@thefnmark}$\hss}}    
        
\def\ps@myheadings{\let\@mkboth\@gobbletwo
\def\@oddhead{\hbox{}
\rightmark\hfil\eightrm\thepage}   
\def\@oddfoot{}\def\@evenhead{\eightrm\thepage\hfil
\leftmark\hbox{}}\def\@evenfoot{}
\def\sectionmark##1{}\def\subsectionmark##1{}}

\oddsidemargin=\evensidemargin
\renewcommand{\thefootnote}{\fnsymbol{footnote}}

\newcounter{sectionc}\newcounter{subsectionc}\newcounter{subsubsectionc}
\renewcommand{\section}[1] {\vspace{12pt}\addtocounter{sectionc}{1} 
\setcounter{subsectionc}{0}\setcounter{subsubsectionc}{0}\noindent 
        {\tenbf\thesectionc. #1}\par\vspace{5pt}}
\renewcommand{\subsection}[1] {\vspace{12pt}\addtocounter{subsectionc}{1} 
        \setcounter{subsubsectionc}{0}\noindent 
        {\bf\thesectionc.\thesubsectionc. {\kern1pt \bfit #1}}\par\vspace{5pt}}
\renewcommand{\subsubsection}[1] {\vspace{12pt}\addtocounter{subsubsectionc}{1}
        \noindent{\tenrm\thesectionc.\thesubsectionc.\thesubsubsectionc.
        {\kern1pt \tenit #1}}\par\vspace{5pt}}
\newcommand{\nonumsection}[1] {\vspace{12pt}\noindent{\tenbf #1}
        \par\vspace{5pt}}

\newcounter{appendixc}
\newcounter{subappendixc}[appendixc]
\newcounter{subsubappendixc}[subappendixc]
\renewcommand{\thesubappendixc}{\Alph{appendixc}.\arabic{subappendixc}}
\renewcommand{\thesubsubappendixc}
        {\Alph{appendixc}.\arabic{subappendixc}.\arabic{subsubappendixc}}

\renewcommand{\appendix}[1] {\vspace{12pt}
        \refstepcounter{appendixc}
        \setcounter{figure}{0}
        \setcounter{table}{0}
        \setcounter{lemma}{0}
        \setcounter{theorem}{0}
        \setcounter{corollary}{0}
        \setcounter{definition}{0}
        \setcounter{equation}{0}
        \renewcommand{\thefigure}{\Alph{appendixc}.\arabic{figure}}
        \renewcommand{\thetable}{\Alph{appendixc}.\arabic{table}}
        \renewcommand{\theappendixc}{\Alph{appendixc}}
        \renewcommand{\thelemma}{\Alph{appendixc}.\arabic{lemma}}
        \renewcommand{\thetheorem}{\Alph{appendixc}.\arabic{theorem}}
        \renewcommand{\thedefinition}{\Alph{appendixc}.\arabic{definition}}
        \renewcommand{\thecorollary}{\Alph{appendixc}.\arabic{corollary}}
        \renewcommand{\theequation}{\Alph{appendixc}.\arabic{equation}}
        \noindent{\tenbf Appendix \theappendixc #1}\par\vspace{5pt}}
\newcommand{\subappendix}[1] {\vspace{12pt}
        \refstepcounter{subappendixc}
        \noindent{\bf Appendix \thesubappendixc. {\kern1pt \bfit #1}}
        \par\vspace{5pt}}
\newcommand{\subsubappendix}[1] {\vspace{12pt}
        \refstepcounter{subsubappendixc}
        \noindent{\rm Appendix \thesubsubappendixc. {\kern1pt \tenit #1}}
        \par\vspace{5pt}}

\newcommand{\textlineskip}{\baselineskip=13pt}
\newcommand{\smalllineskip}{\baselineskip=10pt}

\def\eightcirc{
\begin{picture}(0,0)
\put(4.4,1.8){\circle{6.5}}
\end{picture}}
\def\eightcopyright{\eightcirc\kern2.7pt\hbox{\eightrm c}} 

\newcommand{\copyrightheading}[1]
        {\vspace*{-2.5cm}\smalllineskip{\flushleft
        {\footnotesize International Journal of Modern Physics A, #1}\\
        {\footnotesize $\eightcopyright$\, World Scientific Publishing
         Company}\\
         }}

\def\abstracts#1#2#3{{
        \centering{\begin{minipage}{4.5in}\baselineskip=10pt\footnotesize
        \parindent=0pt #1\par 
        \parindent=15pt #2\par
        \parindent=15pt #3
        \end{minipage}}\par}} 

\renewenvironment{thebibliography}[1]
        {\frenchspacing
         \ninerm\baselineskip=11pt
         \begin{list}{\arabic{enumi}.}
        {\usecounter{enumi}\setlength{\parsep}{0pt}
         \setlength{\leftmargin 12.7pt}{\rightmargin 0pt} 
         \setlength{\itemsep}{0pt} \settowidth
        {\labelwidth}{#1.}\sloppy}}{\end{list}}

\newcounter{itemlistc}
\newcounter{romanlistc}
\newcounter{alphlistc}
\newcounter{arabiclistc}

\newcommand{\fcaption}[1]{
        \refstepcounter{figure}
        \setbox\@tempboxa = \hbox{\footnotesize Fig.~\thefigure. #1}
        \ifdim \wd\@tempboxa > 5in
           {\begin{center}
        \parbox{5in}{\footnotesize\smalllineskip Fig.~\thefigure. #1}
            \end{center}}
        \else
             {\begin{center}
             {\footnotesize Fig.~\thefigure. #1}
              \end{center}}
        \fi}

\newcommand{\tcaption}[1]{
        \refstepcounter{table}
        \setbox\@tempboxa = \hbox{\footnotesize Table~\thetable. #1}
        \ifdim \wd\@tempboxa > 5in
           {\begin{center}
        \parbox{5in}{\footnotesize\smalllineskip Table~\thetable. #1}
            \end{center}}
        \else
             {\begin{center}
             {\footnotesize Table~\thetable. #1}
              \end{center}}
        \fi}

\def\@citex[#1]#2{\if@filesw\immediate\write\@auxout
        {\string\citation{#2}}\fi
\def\@citea{}\@cite{\@for\@citeb:=#2\do
        {\@citea\def\@citea{,}\@ifundefined
        {b@\@citeb}{{\bf ?}\@warning
        {Citation `\@citeb' on page \thepage \space undefined}}
        {\csname b@\@citeb\endcsname}}}{#1}}

\newif\if@cghi
\def\cite{\@cghitrue\@ifnextchar [{\@tempswatrue
        \@citex}{\@tempswafalse\@citex[]}}
\def\citelow{\@cghifalse\@ifnextchar [{\@tempswatrue
        \@citex}{\@tempswafalse\@citex[]}}
\def\@cite#1#2{{$\null^{#1}$\if@tempswa\typeout
        {IJCGA warning: optional citation argument 
        ignored: `#2'} \fi}}

\def\pmb#1{\setbox0=\hbox{#1}
        \kern-.025em\copy0\kern-\wd0
        \kern.05em\copy0\kern-\wd0
        \kern-.025em\raise.0433em\box0}

\def\fnt#1#2{\footnotetext{\kern-.3em
        {$^{\mbox{\scriptsize #1}}$}{#2}}}

\def\fpage#1{\begingroup
\voffset=.3in
\thispagestyle{empty}\begin{table}[b]\centerline{\footnotesize #1}
        \end{table}\endgroup}

\def\runninghead#1#2{\pagestyle{myheadings}
\markboth{{\protect\footnotesize\it{\quad #1}}\hfill}
{\hfill{\protect\footnotesize\it{#2\quad}}}}
\font\tenrm=cmr10
\font\tenit=cmti10 
\font\tenbf=cmbx10
\font\bfit=cmbxti10 at 10pt
\font\ninerm=cmr9

\font\eightrm=cmr8

\def\qed{\hbox{${\vcenter{\vbox{                        
   \hrule height 0.4pt\hbox{\vrule width 0.4pt height 6pt
   \kern5pt\vrule width 0.4pt}\hrule height 0.4pt}}}$}}

\renewcommand{\thefootnote}{\fnsymbol{footnote}}        

\begin{document}

\runninghead{C. V. Johnson, ``The Enhan\c con,
  Multimonopoles$\ldots$''} {C. V. Johnson, ``The Enhan\c con,
  Multimonopoles$\ldots$''}

\normalsize\textlineskip
\thispagestyle{empty}
\setcounter{page}{1}

\copyrightheading{}                     

\vspace*{0.88truein}

\fpage{1} \centerline{\bf THE ENHAN\c CON, MULTIMONOPOLES AND FUZZY
  GEOMETRY} \vspace*{0.3truein} \centerline{\footnotesize CLIFFORD V.
  JOHNSON\footnote{\tt c.v.johnson@durham.ac.uk}} \vspace*{0.3truein}
\centerline{\footnotesize\it Department of Mathematical Sciences}
\baselineskip=10pt \centerline{\footnotesize\it University of Durham,
  South Road,} \baselineskip=10pt \centerline{\footnotesize\it Durham
  DH1 3LE, United Kingdom}

\date{16th October 2000}

\vspace*{0.21truein} \abstracts{The presentation at Strings 2000 was
  intended to be in two main parts, but there was only time for part
  one. However both parts appeared on the online proceedings, and are
  also included in this document. The first part concerns an
  exploration of the connection between the physics of the ``enhan\c
  con'' geometry arising from wrapping $N$ D6--branes on the K3
  manifold in Type~IIA string theory and that of a charge $N$ BPS
  multi--monopole. This also relates to the physics of 2+1 dimensional
  $SU(N)$ gauge theory with eight supercharges. The main results
  uncovered by this exploration are: {\it a)} better insight into the
  non--perturbative geometry of the enhan\c con; {\it b)} the
  structure of the moduli space geometry, and its characterisation in
  terms of generalisations of an Atiyah--Hitchin--like manifold; {\it
    c)} the use of Nahm data to describe aspects of the geometry, showing
  that the enhan\c con locus itself has a description as a fuzzy
  sphere.  Part two discusses the addition of extra D2--branes into
  the geometry.  Two probe computations show the difference between
  the geometry as seen by D2--branes and that seen by wrapped
  D6--branes, and the accompanying gauge theory interpretations are
  discussed.}{}{}

\textheight=7.8truein
\setcounter{footnote}{0}
\renewcommand{\thefootnote}{\alph{footnote}}

\section{Wrapping Branes}
\noindent

In the context of superstring theory, one of the simplest ways to
realize a supersymmetric $SU(N)$ gauge theory with no flavours and
only eight supercharges is {\it via} the low energy limit of $N$
D--branes wrapped on manifolds which break half of the supersymmetry
of the parent string theory. The brane breaks the other half.

For $D=4$, ${\cal N}=2$, or $D=3$, ${\cal N}=4$, the appropriate thing
to do is wrap D7--branes or (respectively) D6--branes, on the manifold
$K3$. Let us focus on the latter. We would like to understand as much
as we can about the gauge theory by studying the physics of these
wrapped branes. Perhaps new insights can be gained at large
$N$, since in that case the back reaction on the supergravity fields
is large, and some of the physics of the branes ---and hence the gauge
theory --- is encoded in a non--trivial supergravity solution.

The characteristic scale of the curvature of the supergravity solution
is $~L=~g_s \ell_s N$, (where $g_s$ is the string coupling and
$\ell_s^2~=~\alpha^\prime~=~1/(2\pi T)$, with $T$ the string tension)
while the $SU(N)$ Yang--Mills coupling is:
\begin{equation}
g_{\rm YM}^2=(2\pi)^4 \ell_s^3 g_s V^{-1}\ .
\label{coupling}
\end{equation}
Here, $V$ is the volume of the $K3$ upon which we wrap the branes.

With $N$ large we can stay at weak string coupling and low energy
$(g_s,\ell_s~\ll~1)$, while sensibly discussing supergravity (without
quantum and stringy effects), since~$L$ can be large. Asking that $V$
be small enough (for a suitable wrapping to the resulting 2+1
dimensional theory) keeps us in the infra--red (IR) of the gauge
theory, where $g^{\phantom{2}}_{\rm YM}$ is large. Therefore, if we
are going to learn anything at all from the gravity solution, it will
concern the gauge theory at both large $N$ and strong coupling.

While all of the above is true, it does not guarantee that we will
succeed in our quest, since $a)$ We need to make sure that we find a
smooth and sensible supergravity geometry and $b)$ We must understand
it well enough to deduce anything from it, and $c)$ There is no
guarantee that there is necessarily a clean separation between the
supergravity physics and that of the gauge theory in a way which
achieves a complete duality between the two as in the case with the
AdS/CFT correspondence.\cite{juan}

Issues $a)$ and $b)$ were addressed in ref.[2], 
where many aspects of the solution were uncovered. In particular, the
main naive geometry one would write for the $K3$ wrapped D6 is
singular, and ref.[2]
showed how this naive geometry
(containing a ``repulson'' singularity\cite{repulsive}) is replaced by
a non--singular one. Essentially, the correct geometry contains a
spherical shell of smeared D6--branes which has radius
\begin{equation}
r_{\rm e}={V^*\over V-V^*} g_s N \ell_s\ ;
\label{radius}
\end{equation}
where $V^*\equiv (2\pi\ell_s)^4$ is a critical volume which we shall
discuss later.\footnote{Note that versions of this story for other
  gauge groups have been worked out in ref.[4]
}

Issue $c)$ is a subtle one which requires further work. As pointed out
in ref.[2], 
there is {\it no} clear separation between supergravity and the gauge
theory, and so either the only recourse is to study the entire type
IIA string theory in this background, { or} to seek some other
effective theory in the limit, which might capture the full dynamics
at large $N$. Some conjectural remarks in the latter direction are
contained in ref.[3] 
and we will not pursue this further
here.

Let us instead focus on learning more about the low energy dynamics of
the gauge theory and large $N$, and attempt to see what we can learn.

\section{Probe Results}
The supergravity solution for the $N$ wrapped D6--branes is naively:
\begin{eqnarray}
ds^2=(Z_2Z_6)^{-\frac{1}{2}}(-dt^2+dx^2_1+dx^2_2)
+(Z_2Z_6)^{\frac{1}{2}}(dr^2+r^2d\Omega_2^2)
+\left(V\frac{Z_2}{Z_6}\right)^{\frac{1}{2}}ds^2_{K3}\ ,
\label{thesolution}
\end{eqnarray}
with $ds^2_{K3}$ the metric of $K3$, normalised to unit volume, and
the harmonic functions are
\begin{eqnarray}
Z_2 &=& 1-\frac{ (2\pi)^4 g_s N \alpha'^{5/2} }{ 2Vr } \ , \quad
Z_6 = 1+\frac{g_sN\alpha'^{1/2}}{2r} \ ,
\label{harmonic}
\end{eqnarray}
and ${d\Omega}_2^2={d\theta}^2+\sin^2\!\theta\, {d\phi}^2$.  The true
volume of $K3$ varies with radius and is $V(r)=V Z_2(r)/Z_6(r)$. The
repulson singularity is at $V(r)=0$.

One of the telling results of the analysis of ref.[2]
was the result that the metric seen by a single wrapped D6--brane
probe in the fields produced by all the others is given by:
\begin{equation}
ds^2=
F(r) \left({d r}^2
 +r^2{d\Omega}_2^2 \right)
+F(r)^{-1}\left({d s}/2-\mu_2C_\phi{d\phi}/2\right)^2\ ,
\label{fourth}
\end{equation}
where
\begin{equation}
F(r)={Z_6\over 2g_s}\left(\mu_6V(r)-\mu_2\right)\ ,
\end{equation}
and $C_\phi={-}(r_6/g_s)\cos\theta$.  The basic D6-- and D2--brane
charges are\cite{polchinski}
$\mu_6{=}(2\pi)^{-6}\alpha^{\prime{-7/2}}$ and
$\mu_2{=}(2\pi)^{-2}\alpha^{\prime -3/2}$,
respectively.\footnote{Strictly speaking, the spirit of our discussion
  says that we should have replaced $N$ by $N-1$ in the probe result,
  since we have separated off the probe. This is not really essential,
  and so we will not do it.}

This moduli space metric shows that the kinetic term of the brane
vanishes at the enhan\c con radius ${r}_{\rm e}$, and so it cannot
proceed inwards of that radius.  So in equations (\ref{thesolution}),
the geometry for $r<{ r}_{\rm e}$ cannot be correct, since it is that
associated with branes at $r=0$.  The idea of ref.[2],
bolstered by other facts we will mention shortly, ({\it e.g} $V(r_{\rm
  e})=V^*$) is that to a first approximation, the geometry in the
interior should be a flat, sourceless geometry, matched onto the
exterior solution in eqn.~(\ref{thesolution}) at $r={ r}_{\rm
  e}$.\footnote{See ref.[7]
  for a supergravity discussion of the physics of this sort of
  excision process.}

The limit where we isolate the massless fields ---looking for the
gauge theory--- is to take $\alpha^\prime{\to}0$, holding the gauge
coupling (\ref{coupling}) and $U{=}r/\alpha'$ finite.\cite{juan} (The
latter should be thought of as the characteristic energy scale at
which the $SU(N)$ breaks to $SU(N-1)\times U(1)$ representing the
separating of the probe.)  In this case, the metric becomes
\begin{eqnarray}
&&ds^2=f(U) \left({d U}^2 +U^2{d\Omega}^2_2\right) +f(U)^{-1}
 \left({d\sigma} -{N\over{8\pi^2}}A_\phi{d\phi}\right)^2\ ,
\nonumber\\
{\rm where}&&\nonumber\\
&&f(U)={1\over 8\pi^2 g^2_{\rm YM}}
\left(1-{ g^2_{\rm YM}N \over U}\right)\ ,
\label{probe}
\end{eqnarray}
the $U(1)$ monopole potential is $A_\phi=-\cos\theta,$ and
$\sigma=s{\alpha^\prime}/2$.  This metric is meaningful only for
$U{>}g^2_{\rm YM}N$. It is the Euclidean Taub--NUT metric, with a
negative mass.  It is a hyperK\"ahler manifold, because $\nabla
f{=}\nabla{\times}A$, where $A{=}(N/8\pi^2)A_\phi d\phi$.

There are many facts about the geometry and the gauge theory which we
can understand in this metric. First, for geometry, the radius $U_{\rm
  e}=g^2_{\rm YM}N$ is the place where the spacetime should be treated
carefully. In fact, it is the scaled version of the ``enhan\c con''
radius $r_{\rm e}$, (see eqn.(\ref{radius})).  Here the volume of the
$K3$ becomes equal to $V^*$, and a generic $U(1)$ in the six
dimensional supergravity is joined by massless W--bosons to acheive an
enhanced $SU(2)$ gauge theory.  The wrapped branes are BPS monopoles
of that $U(1)$, and as such cease to be pointlike and smear out at
symmetry enhancement.  Thus the geometry of spacetime is different
from that which one would write for the case when the wrapped branes
are pointlike.

\begin{figure}[ht]
  \centerline{\psfig{figure=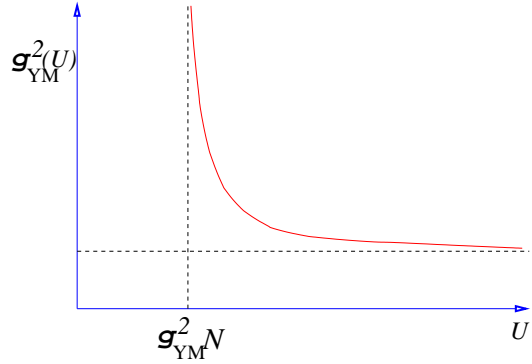,height=2.0in}}
\caption{\small The running gauge coupling.}
\label{running}
\end{figure}

Concerning gauge theory, the factor $f(U)$
represents the one loop running of the Yang--Mills coupling (see
figure \ref{running}).  This is because $U$ is the vacuum expectation
value (vev) of the complex scalar $\phi$ in the ${\cal N}=2$ gauge
multiplet which breaks $SU(N)\to SU(N-1)\times U(1)$. In the low energy
effective theory, this vev controls the effective coupling, since the
model becomes a sigma model whose kinetic term is
$f(U)\partial_\mu U\partial^\mu U$.

The coupling diverges at the scale $U=g^2_{\rm YM}N$. This should be
the scale at which the $SU(N)$ is restored, but in fact it is not,
since the divergence happens at $U\neq 0$. In other words, the branes
are not coincident, and so (extending the discussion to any of the
constituent branes), the generic gauge group is $U(1)^{N-1}$: we are
out on the Coulomb branch of the gauge theory, and the origin is
modified from being an $SU(N)$ point by quantum corrections.

\section{Moduli Space Metrics}

What we looked at just now was a four dimensional submanifold of the
full $4N-4$ dimensional moduli space of the $SU(N)$ gauge theory. This
full moduli space, which is hyperK\"ahler, is proposed to be
isomorphic to the relative moduli space\cite{modulispace} of $N$ BPS
monopoles of 4D $SU(2)$ gauge theory spontaneously broken to
$U(1)$.\cite{SWtwo,CH,hanany} Here, the wrapped D6--branes play the
role of the monopoles, since they couple magnetically to the $U(1)$
which gets enhanced to an $SU(2)$ when $V(r)=V^*$. (There are 23 other
$U(1)$'s which will not concern us here.)  In the directions
perpendicular to the wrapped brane, with coordinates
$(r,\theta,\phi)$, they appear as monopoles in the $3+1$ dimensional
problem.

So let us learn both about the gauge theory and the monopole moduli
space. First let us consider the case of two monopoles $(N=2)$. In that
case, it is known that the metric on the gauge theory moduli space is
the Atiyah--Hitchin manifold, (as for the two monopole moduli space) which
is given as follows:\cite{atiyah,gibbonsmanton}
\begin{eqnarray}
&&ds^2_{\rm AH}=f^2d\rho^2+a^2\sigma_1^2+
b^2\sigma_2^2+c^2\sigma_3^2\ ,\,\,\,{\rm where}\nonumber \\
&&\qquad\sigma_1=-\sin\psi d\theta+\cos\psi\sin\theta d\phi\ ;\nonumber\\
&&\qquad\sigma_2=\cos\psi d\theta+\sin\psi\sin\theta d\phi\ ;\nonumber\\
&&\qquad\sigma_3=d\psi +\cos\theta d\phi\ ;\nonumber\\
{2bc\over f}{da\over d\rho}&=&(b-c)^2-a^2\ ,\mbox{ and cyclic perms.;}\quad 
\rho =2K\left(k\right),
\end{eqnarray} and $K(k)$ is the elliptic integral of the first kind:
\begin{equation}
K(k)=\int_0^{\pi\over2}(1-k^2\sin^2\tau)^{1\over2}d\tau\ .
\label{elliptic}
\end{equation}
Also, $k$, the ``modulus'', runs from $0$ to $1$, so
$\pi\leq\rho\leq\infty$.

In our moduli space metric (\ref{probe}), if we rescale
$\rho=U/g^2_{\rm YM}$ and $\psi=4\pi^2\sigma$, then we get:
\begin{eqnarray}
&&ds^2= {g^2_{\rm YM}\over 8\pi^2}ds^2_{\rm TN}\ ,
\quad{\rm with}\label{taubnut}\\
&&ds^2_{\rm TN}=\left(1-{2\over\rho}\right) \left({d \rho}^2 
+\rho^2{d\Omega}^2_2\right) 
+4\left(1-{2\over\rho}\right)^{-1}
\left({d\psi}+\cos\theta{d\phi}\right)^2\ .
\nonumber
\label{negative}
\end{eqnarray}
This is precisely the form of the metric that one gets by expanding
the Atiyah--Hitchin metric in large~$\rho$. The difference between the
Atiyah--Hitchin manifold and this ``negative mass'' ($m=-1$ in
standard units, see {\it e.g.} ref.[15])
Taub--NUT metric is a family of exponential corrections of the form
$e^{-\rho}$, which are non--perturbative in a large $\rho$ expansion.

{}From the point of view of the gauge theory, these corrections are of
order $e^{-U/g^2_{\rm YM}}$, and so are instanton corrections. These
corrections render the complete manifold smooth ---the $\rho=2$
singularity is just an artifact of the large $\rho$ expansion--- which
is consistent with gauge theory expectations.

The combination $\psi=4\pi^2\sigma={\tilde \sigma}$ is the dual scalar
to the gauge theory photon and has period $2\pi$. The fact that $\psi$
is $2\pi$ periodic and not $4\pi$ gives us an $SO(3)$ isometry for the
Atiyah--Hitchin manifold instead of the naive $SU(2)$ of the Taub--NUT
metric. In fact, it can be proven that the {\it unique} smooth
hyperK\"ahler manifold with this isometry and the above asymptotic
behaviour is the Atiyah--Hitchin manifold, essentially proving that it
is the correct result for the $SU(2)$ gauge theory moduli
space.\cite{SWtwo,valya}

What about $N$ different from 2? Well, the key observation\cite{fuzzy}
is that the rescalings $\rho=2U/g^2_{\rm YM}N$ and $\psi=\sigma
8\pi^2/N$, give
\begin{equation}
ds^2= {g^2_{\rm YM}N^2\over 32\pi^2}ds^2_{\rm TN}\ ,
\end{equation}
generalising our previous result (\ref{negative}) connecting the
moduli space metric to the same negative mass Taub--NUT spacetime. It
is tempting to conclude that the manifold controlling our
non--perturbative corrections for all $N$ is again the Atiyah--Hitchin
manifold, but we should be careful\footnote{We confess to being less
  than careful in the original version of ref.[3], 
  and also in the Strings 2000 talk and the transparencies displayed
  at {\tt http://feynman.physics.lsa.umich.edu/strings2000/}}.  Note
that changing $N$ will not change the period of the probe $U(1)$.
${\tilde \sigma}=4\pi^2\sigma$ should still have period $2\pi$.
Therefore the period of $\psi$ in the metric above is now $4\pi/N$.
Furthermore, the instanton corrections in the gauge theory will still
be typically $e^{-U/g^2_{\rm YM}}$, and so this translates into
$e^{-N\rho/2}$ for the corrections to the $m=-1$ Taub--NUT metric.

\medskip

Let us then summarise what we have learned for our non--perturbative
metrics for any $N$. They are Atiyah--Hitchin--like in the sense that:

$a)$ They differ from negative mass Taub--NUT by non--perturbative
corrections of the form $e^{-N\rho/2}$.

$b)$ They have only a {\it local} $SU(2)$ action. Globally, it is
broken by instanton corrections to $SU(2)/{\mathbf Z}_N$.

\medskip

Some remarks are in order. One might worry that the manifolds that we
seek might be singular, especially given property $b)$. This would be
an awkward feature from the gauge theory perspective, since there is
nothing like a Higgs branch available which might be opening up at the
singularity, which is typically what happens in situations like this.
(Classically there is no Higgs branch, and it seems imprudent to
conjecture that one might be generated quantum mechanically.)

However, it is natural that there might be a harmless singularity in
the moduli space following from the fact that we are studying a four
dimensional subspace of the full $4N-4$ dimensional moduli space,
which is known to be smooth. It is natural to expect that there are
therefore $4N-8$ adjustable parameters in the full metric, which can
resolve the singularity. The worry that the parameters do not appear
in the metric written above is alleviated by the fact that these are
non--perturbative parameters. They are monopole positions and phases:
parameters of different vacua from the point of view of the one we
have focussed on.

{}From the point of view of the monopole problem, or from the enhan\c
con picture, this simply corresponds to the other monopoles which make
up the internal structure of the $N-1$ monopole having to move around
somewhat in order to accommodate the merging with the single monopole
probe.

Property $a)$ also deserves some remarks. While the instanton
corrections thus identified will smooth the singularity (together with
the possible need for a resolution by adjusting $4N-8$ parameters) in
the moduli space, it is clear that in the strict large $N$ limit, the
corrections are extremely small because of the $N$ in the exponential.
So in fact, the strict supergravity result for the location and nature
of the enhan\c con is quite accurate. (See also ref.[20])

It is worth comparing this to the two monopole case ($N=2$). There,
the naive singularity of the Taub--NUT metric is at $\rho=2$. It is in
fact completely removed by non--perturbative effects, and the radius
$\rho=\pi$ is the smallest possible value. This corresponds to the
axisymmetric case of the two monopoles being exactly on top of one
another.\cite{ward} (This is naively a singularity of the complete
metric but is in fact a ``bolt'' singularity, which is harmless.)

It is clear that there is a natural generalisation of this for
each~$N$. For very large~$N$, the coincident monopole case is the
enhan\c con itself, at $\rho=2$ of $U=g^2_{\rm YM}N$.

Another point worth making about large $N$ here is that while $\rho$
is a relative coordinate in the case of the two--monopole case, and
the Taub--NUT and Atiyah--Hitchin manifolds have nothing to do with
spacetime geometry, this is not true here: For large $N$, $\rho$ is a
rescaled centre of mass coordinate of the $N-1$ monopole system being
probed by a single one. But since $N$ is large, the centre of mass is
essentially at the centre of the $N-1$ monopole, and so serves as a
good spacetime coordinate, with the large monopole at the origin.


\section{Fuzzy Geometry}
\label{fuzzed}
\begin{figure}[ht]
  \centerline{\psfig{figure=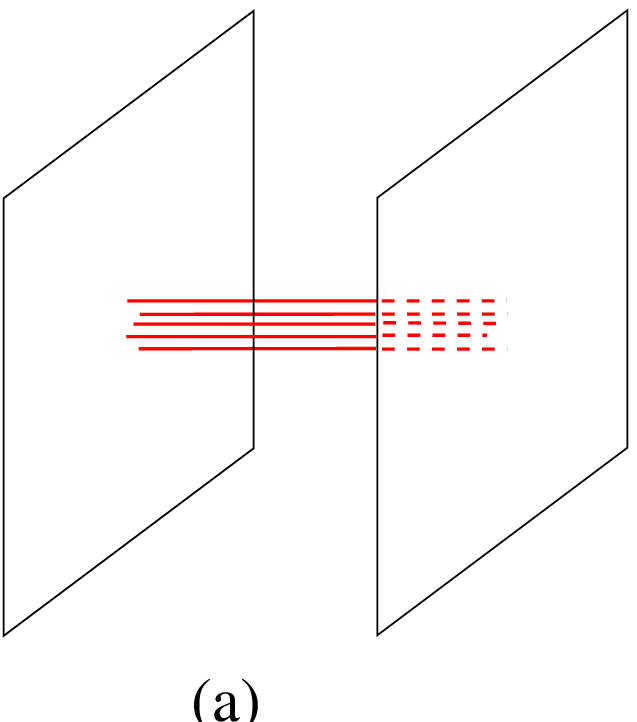,height=2.0in}
    \hskip2cm\psfig{figure=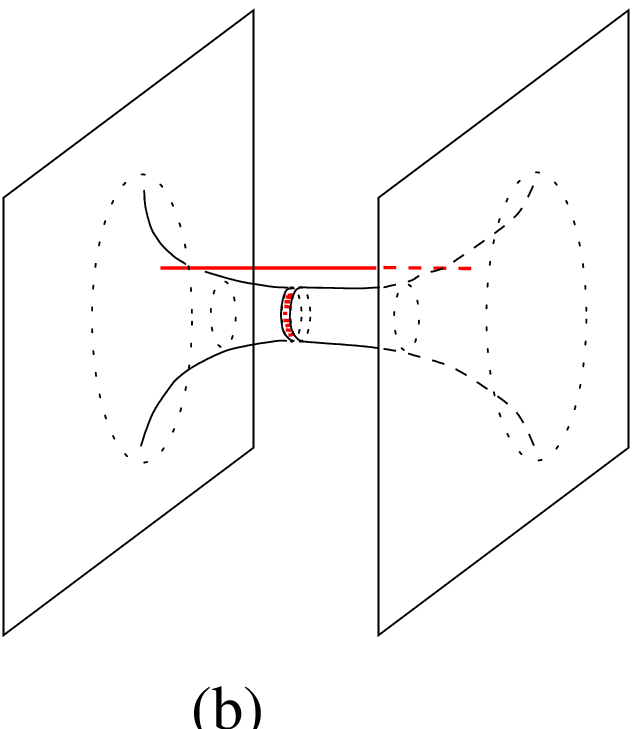,height=2.1in}}
\caption{\small
  (a) A slice of the configuration of D3--branes stretching between
  NS5--branes.  (b) A slice of the resulting ``double trumpet'' shape
  of the NS5--branes at large~$N$. (A separated probe is also shown.)
  This system has a natural description in terms of the Nahm equations
  as explained in the text. The enhan\c con is the place (an $S^2$)
  where the NS5--branes touch.}
\label{trumpet}
\end{figure}

Another appearance of the enhan\c con geometry is when a large number,
$N$, of D3--branes are stretched between a pair of NS5--branes, as in
figure~\ref{trumpet}(a).
This situation is related to the $K3$--wrapped D6--brane system by a
chain of dualities.\cite{jpp} The D3--branes pull the NS5--branes into
a shape such that at large $N$ they meet in an $S^2$ geometry, as in
figure~\ref{trumpet}(b). This $S^2$ is the enhan\c con, since when two
NS5--branes in type~IIB string theory meet, there is an enhanced
$SU(2)$ gauge symmetry describing their relative collective geometry.
The endpoints of the D3--branes act as monopoles of the spontaneously
broken $SU(2)$, where the Higgs vev is the asymptotic distance between
the NS5--branes.

It is intuitively clear that in this and the wrapped D6--brane
picture, the enhan\c con geometry is only a smooth sphere at large
$N$. The geometry is made of $N$ smeared objects, the monopoles.

To find a quantitative description of this geometry needs new
variables. In this D3/NS5 system, the variables are intuitive: $N$
D3--branes have 3 $N\times N$ matrix valued coordinates, $\Phi^i$
($i=1,2,3$), representing their transverse  motions within the
NS5--brane worldvolumes. These coordinates are  governed by Nahm's
equations in the BPS case:\cite{nahm,diaconescu}
\begin{equation}
{d\Phi^i\over d\sigma}=
{1\over 2}\epsilon_{ijk}[\Phi^j,\Phi^k]\ ,
\label{nahm}
\end{equation}
where $\sigma\in[-1,1]$, (where we have placed the NS5--branes at
$\pm1$) and the appropriate solutions are those for which the
$\Phi^i(\sigma)$ have a pole at the ends of the interval, with
residues $\Sigma^i$, which form the $N\times N$ irreducible
representation of $SU(2)$:\cite{hitchin,nakajima,tsimpis}
\begin{equation}
[\Sigma^i,\Sigma^j]=2{\rm i}\epsilon^{ijk}\Sigma^k\ .
\end{equation}
It is clear from this description that the D3--brane geometry has a
non--commutative geometry underlying its description, the general
solution taking one $SU(2)$ representation and twisting it into
another while traversing the interval. A slice through this geometry,
such as the enhan\c con itself at $\sigma=0$, is clearly a
non--commutative or ``fuzzy'' sphere.\cite{fuzzysphere} As a simple
attempt at creating an approximate solution, a symmetric ``double
trumpet'' configuration can be made by gluing two sections of the
infinite funnel solution of ref.[21]:
\begin{equation}
\Phi^i=f(\sigma)\Sigma^i\ ,\qquad f(\sigma)={g_s\over \sigma\pm1}\ . 
\end{equation}
A complete solution would of course connect smoothly through the
middle, but this crude solution serves to illustrate the
point\footnote{The idea is to make a solution which is spherically
  symmetric. The only monopole solution which is spherically symmetric
  is the one--monopole.\cite{onemono} But we expect that for large
  $N$, we can make a solution which approaches spherical symmetry on
  the sufficiently granular (supergravity) scale.}: A cross section of
this is a fuzzy sphere, with radius $R$ given by:
\begin{equation}
R^2=4\pi^2\alpha^\prime \sum_{i=1}^3{\rm
  Tr}(\Phi^i)^2=4\pi^2\alpha^\prime (N^2-1)f(\sigma)\ ,
\end{equation}
giving a radius for the enhan\c con (at $\sigma=0$) of $R_{\rm e}\sim
g_s \ell_s N$, which is we saw in the wrapped D6--brane description.

It is noteworthy that the geometry of this object made of expanding
branes is described as a fuzzy sphere just as the brane expansion in a
background R--R field produced by the dielectric brane
effect.\cite{robdielectric} One is led to speculate that there might
be further couplings to be deduced, such as multipole couplings to
{\it higher} rank R--R fields {\it via} curvature terms on the
D--brane worldvolume.\footnote{Contrast this with the route taken in
  ref.[2], 
  where it was deduced indirectly that the branes smear, using the
  electric couplings\cite{vafa} to {\it lower} rank R--R fields,
  producing a place where the tension will go to zero, W--bosons of an
  enhanced gauge symmetry, implying BPS monopoles, smearing, {\it etc.}}

\newpage

\section{Other Branes and other Probes}

Now although the wrapping of a D6--brane on $K3$  induces a negative
D2--brane charge within the part of its worldvolume transverse to
$K3$, it must be stressed that this is not an anti--D2--brane. It
preserves the same supersymmetries that a D2--brane with that
orientation would.

In fact, it is useful to consider adding extra D2--branes to the
wrapped D6--brane arrangement with the same orientation as the
unwrapped part of the D6--branes. Let us add $M$ of them. The net
D2--brane charge as measured at infinity is therefore $M-N$, and it is
interesting to note that we can cancel away the net charge by setting
$M=N$. 

There is an $SU(M)$ gauge theory on the worldvolume of these new
branes, and there are of course strings which can stretch between
these branes and the other set. Since these ones are not wrapped, the
2+1 dimensional gauge theory coupling is simply
\begin{equation}
g^2_{\rm YM,2}=g_s/\ell_s\ .
\label{newcoupling}
\end{equation}
This form will produce an interesting conflict with the gauge theory
coupling (\ref{coupling}) of the wrapped brane gauge theory when we
take decoupling limits shortly.

A natural question to ask is what happens to the geometry with the
addition of these extra branes? Well, the first thing to note is that
the supergravity geometry is simply modified by changing the harmonic
function (\ref{harmonic}) which yields the D2--brane charge:
\begin{equation}
Z_2 =1 -\frac{ (2\pi)^4 g_s N \alpha'^{5/2} }{ 2Vr } \longrightarrow
{\widetilde Z}_2= 1 +\frac{ (2\pi)^4 g_s (M{-}N) \alpha'^{5/2} }{ 2Vr } \ .
\end{equation}
Note that this modifies, for example, the result for the volume of
$K3$, which is now $V(r)=V{\widetilde Z}_2/Z_6$.

The next thing to notice is that we can probe with wrapped D6--branes
as before, but we also can probe with pure D2--branes. The result for
probing with a wrapped D6--brane is much like that which we saw before. It
is equation (\ref{fourth}), but now with
\begin{equation}
F(r)={Z_6\over 2g_s}\left(\mu_6V(r)-\mu_2\right)=
{1\over 2g_s}\left(\mu_6{\widetilde Z}_2-\mu_2Z_6\right)\ .
\label{changed}
\end{equation}
This means that everything we have seen is generically the same as
before. There is an enhan\c con, which is still where $V(r)=V^*$, but
for increasing $M$, this happens at smaller radius, given by:
\begin{equation}
{\tilde r}_{\rm e}={V^*\over  V-V^*}g_s \ell_s (N-M)\ .
\end{equation} 
After taking a decoupling limit, we will get similar results with $N$
replaced by $N-M$, and so, for example the scaled enhan\c con radius
is ${\widetilde U}_{\rm e}=(N{-}M)g^2_{\rm YM}$.  The result for probing
with the pure D2--brane is easy to deduce. Take the previous results
(\ref{fourth}) and (\ref{changed}), set $\mu_6$ to  zero, and
reverse the sign of $\mu_2$.  So we get in this case
\begin{equation}
F(r)={\mu_2Z_6\over 2g_s}\ .
\label{changedtwo}
\end{equation}
So we see that the pure D2--branes completely ignore the D2--brane
component, and we just get the usual result similar to that for
D2--branes probing D6--branes in flat space. The effects of the
wrapping on the transverse space simply do not show up. This is seen
more clearly when one places this result for $F(r)$ into the metric
(\ref{fourth}) and see that it is essentially the Taub--NUT metric,
now with a positive mass set by $N$. This will be precisely the case
when we take a decoupling limit shortly.

So in fact the $M$ D2--branes which make up the geometry can clearly live
at $r=0$, since there is no reason for them to be stuck at the enhan\c
con radius. After excising the geometry interior to the radius $r_{\rm
  e}$, instead of replacing it with a flat geometry as we did
before,\cite{jpp} we should replace it with a geometry produced by $M$
D2--branes in the interior, sitting at $r=0$, appropriately matched at
${\tilde r}_{\rm e}$ to the exterior geometry.\footnote{See
  ref.[30] 
for further discussion of this sort of procedure.}

This is all made clearer when we take the decoupling limit of this
result. Again we have a choice. We can take $\alpha^\prime\to0$,
holding $U=r/\alpha^\prime$ fixed, but we can choose to hold $a)$ the
Yang--Mills coupling, (\ref{coupling}) of the wrapped D6--branes,
fixed, or $b)$ the coupling (\ref{newcoupling}) on the pure D2--branes
fixed. We cannot do both.  The former stays with the wrapped
D6--brane's $SU(N)$ gauge theory, and treats the D2--branes as merely
supplying extra hypermultiplets\footnote{We must be careful though,
  since the D2--branes are too small to make the gauge dynamics of their
  ``flavour group'', $SU(M)$, decouple. So they are probably only
  formally hypermultiplets.}. The latter limit, instead, treats the
D6--branes as hypermultiplets within the D2--brane $SU(M)$ gauge
theory.

For the case $a)$ we get the metric in equation (\ref{probe})  with
scaled function
\begin{equation}
f(U)={N\over 16\pi^2 U}\ ,
\end{equation}
a pure one loop contribution. Meanwhile, for case $b)$ we have instead
the scaled function
\begin{equation}
  f(U)={1\over 8\pi^2 g^2_{\rm YM,2}}\left(1+{N g^2_{\rm YM,2} \over
      2U}\right)\ .
\end{equation} (It is interesting that one limit allows the tree level
part to survive ---coming from the ``1'' in the metric--- while the other
does not. The second case is reminiscent of an AdS--like limit.) The
latter is precisely what one should get for the moduli space result.
There is a $U(1)$ on the D2--brane and $N/2$ massless flavours supplied
by the D6--branes.

With unwrapped D6--branes, we would have the following to say about
the location $U=0$: The metric is smooth there for one flavour, as the
NUT singularity is removable, but for higher numbers of flavours, the
singularity is not removable (recall that the periodicity of $\sigma$
is fixed by the gauge theory), but has the happy interpretation of
being the origin of a Higgs branch. This is the branch where the
D2--brane becomes a fat instanton inside the D6--brane $SU(N)$ gauge
theory. (More can be found out about this in
refs.[28,29]) 

Here, with wrapped branes we have a subtlety. We have asked that we
excise the naive spacetime geometry inside $r={\widetilde r}_{\rm e}$
(or $U={\widetilde U}_{\rm e}$), and replace it with a geometry which
has no D6--branes inside that radius, so in fact there is no way to
form instantons, and hence no Higgs branch. This appears to fit the
fact that the D6--branes are wrapped on a small $K3$: we would need
${\mathbf R}^4$ to form genuine instantons.

\nonumsection{Acknowledgements}
\noindent
CVJ would like to thank the organisers of the Strings 2000 conference
at the University of Michigan for an invitation to present this work,
and for an extremely enjoyable conference. A comment at the conference
by Alex Buchel and an innocent--seeming question from an anonymous
referee of ref.[3] 
are gratefully acknowledged.

\newpage

\nonumsection{References}

\end{document}